\documentclass{PoS}

\title{Application of Domain Decomposition to the Evaluation of Fermion Determinant
Ratios}

\ShortTitle{Application of Domain Decomposition to the Evaluation of Fermion
Determinant Ratios}

\author{\speaker{Jacob Finkenrath}\thanks{J.F. thanks the Marie Curie ITN STRONGnet for support.}$\;^{,a}$, Francesco Knechtli $^{a}$, Bj\"orn Leder $^{a,b}$
	\\
      $^{a}$ Department of Physics, Bergische Universit\"at Wuppertal\\
      Gaussstr. 20, D-42119 Wuppertal, Germany\\
      $^{b}$ Department of Mathematics, Bergische Universit\"at Wuppertal\\
      Gaussstr. 20, D-42119 Wuppertal, Germany\\
      E-mail: \email{finkenrath@physik.uni-wuppertal.de}}

\abstract{
      We analyze the fluctuations in the case of mass reweighting for $N_f=2$ Wilson fermions.
      We use a domain decomposition factorization of the fermion determinant.
      Ratios of determinants are estimated stochastically.
      We study the stochastic and the ensemble fluctuations as a function of
      the volume $V$ and the mass shift $\Delta m$.
      With our result it is possible to estimate the cost and the effectiveness
      of mass reweighting. 
      In addition we introduce a stochastic estimation for the one flavor case without using
      the square root.
}

\FullConference{The 30th International Symposium on Lattice Field Theory\\
                 June 24 - 29,  2012\\
                 Cairns, Australia}

\begin{document}

\section{Introduction}

It is a well-known challenge to include the fermion determinant into the Boltzmann factor of the desired ensemble.
Many applications, like algorithms with Metropolis acceptance-rejection steps or reweighting methods, require the ratio of such
determinants.
The main problem are the fluctuations of the ratio due to the stochastic and the ensemble noise.
In order to use and to improve such methods it is essential to understand these fluctuations.
In \cite{Finkenrath:2012az} we presented an algorithm where we use the knowledge of these fluctuations to establish the
Partial-Stochastic-Multi-Step-algorithm which reaches a high acceptance rate of $60\%$ up to moderate lattice sizes of $(1.2 \, \rm{fm})^4$.
With this experience and some techniques we used, we study here the fluctuations
in the case of mass reweighting. 

In these proceedings we will analyze the scaling of mass reweighting \cite{Hasenfratz:2008fg} by factorizing
the fluctuations into UV- and IR-dominated terms.
This is done by using domain decomposition \cite{Luscher:2005rx}.
The determinant of the (Wilson-)Dirac operator
is then \mbox{$\det D = \det \hat{D} \det D_{ww} \det D_{bb}$}, where the
Schur complement is given by $\hat{D} = 1 - D^{-1}_{bb} D_{bw} D^{-1}_{ww} D_{wb}$ 
with the Dirac operator in block notation
\begin{equation}
 D= \left[ \begin{array}{cc} D_{bb} & D_{bw} \\ D_{wb} & D_{ww} \end{array} \right] \;
{\rm and \; accordingly \;\;}
 \; D^{-1}= \left[ \begin{array}{cc}
     \mathcal{D}_{bb} & \mathcal{D}_{bw} \\ \mathcal{D}_{wb} & \mathcal{D}_{ww}
     \end{array} \right].
\end{equation}
The operator $D_{bb}$ ($D_{ww}$) is a block-diagonal matrix with the black (white) block Dirac operators on the diagonal.
The Schur complement can be restricted to the support of $D_{wb}$ (using the projector $P$ defined by $D_{wb} P = D_{wb}$) 
without changing its determinant and
its inverse
is then of the form $ \hat{D}^{-1} = 1 - P \mathcal{D}_{bw} D_{wb}$.

\section{Two Flavor Mass Reweighting}

The idea of mass reweighting is to reuse an ensemble which is generated at a specific mass $m_1$ (the ensemble mass)
at a different mass $m_2$ (the target mass). This is possible by correcting the Boltzmann factor of the ensemble \cite{Ferrenberg:1988yz}. The
correction for a configuration $U$ enters as the reweighting factor $W(U,m_1,m_2)$ which is given by
\begin{equation}
 W(U,m_1,m_2) = \frac{ \det D(U,m_2)^{N_f}}{ \det D(U,m_1)^{N_f}} = \frac{1}{\det M^{N_f}}
\end{equation}
with $D(U,m)$ the (Wilson-)Dirac operator, $N_f$ the number of flavors (=2) and the ratio matrix $M=D^{-1}(U,m_2)D(U,m_1)$.
The reweighting factor introduces additional noise in the evaluation of observables
 $\langle\mathrm{O}\rangle_{m_2} =  \frac{\langle\mathrm{O} W\rangle_{m_1}}{\langle W\rangle_{m_1}}$, the ensemble fluctuations.
One can avoid the exact evaluation of the determinant by an unbiased stochastic estimation of the integral
\begin{equation}
\frac{1}{\det M^\dagger M} = \int {\rm D}[\eta] {\rm D}[\eta^\dagger] \exp\left\{-\eta^\dagger M^\dagger M \eta\right\} \longrightarrow \frac{1}{N_{hit}} \sum_{i=1}^{N_{hit}} e^{-\eta_i^\dagger(M^\dagger M - 1)\eta_i}
\end{equation}
where $\eta_i$ are complex Gaussian noise vectors, $N_{hit}$ is the number of the estimates
and one estimation costs one inversion of the Dirac operator.
This introduces stochastic fluctuations which are negligible if and only if the ensemble fluctuations dominate the
statistical error of the measurement.
We analyze these fluctuations by reweighting two $N_f =2$ CLS-ensembles ($\mathrm{O}(a)$ impr. Wilson fermions)
of two different sizes $48\times 24^3$ and $64\times 32^3$
at $\beta = 5.3$ ($a=0.066$ fm) from the pseudoscalar mass of 
$m_{PS} = 440$ MeV to the target mass of $ m_{PS} = 310 $ MeV \cite{Fritzsch:2012wq}.

\subsection{Stochastic Fluctuations}

The variance for $N_{hit}=1$ of the stochastic estimation is given by \cite{Hasenfratz:2002ym}
\begin{equation}
\sigma_{s}^2 = \frac{1}{\det (2 M^\dagger M -1)} - \frac{1}{(\det M^\dagger M)^2}.
\label{eq:vst}
\end{equation}
It follows that the variance is only defined if all eigenvalues of the ratio
matrix $M^\dagger M$ are larger than $1/2$ and that every eigenvalue which
is equal to one produces no stochastic noise.
So every method which shifts the eigenvalues of the ratio matrix
to one improves the estimation.
We will shortly motivate and present two different methods which fulfill this condition, mass interpolation \cite{Hasenfratz:2008fg}
and domain decomposition \cite{Luscher:2005rx}.

It is obvious that the mass reweighting factor $W(U_i,m_1,m_2)$ is known if the spectrum of
the Wilson-Dirac operator $D(U_i,m_2)$ is known
\begin{equation}
  \det M^{-1} = \prod_{i=1}^{12V} \frac{\lambda_{i}(D(m_2))}{\lambda_{i}(D(m_2)) + \Delta m}
  \label{eq:exacW}
\end{equation}
where $\Delta m= m_1 - m_2 $. The product is dominated by the IR-modes. The eigenvalues of 
the ratio matrix $M= 1 + \Delta m \cdot D^{-1}(m_2)$ are given by
\begin{equation}
 \lambda(M) = 1 + \Delta m \cdot \lambda(D^{-1}(m_2)).
  \label{eq:evM}
\end{equation}
For the case that there is no negative eigenvalue 
it follows that $\lambda(M^\dagger M)>1$ (for $\Delta m > 0$). So if we use the Wilson-Dirac operator each eigenvalue produces 
stochastic noise in particular also the UV-modes. If we use the Schur complement, the operator 
$D^{-1}(m_2)$ in Eq. (\ref{eq:evM}) is replaced by an operator
$[\mathcal{D}_{bw}(m_2)-\mathcal{D}_{ww}(m_2) D^{-1}_{ww}(m_1) D_{wb}] D^{-1}_{bb}(m_1) D_{bw} $
which could have eigenvalues with a negative or vanishing real part. With the Schur complement the ratio matrix has
a spectrum which is distributed around one.

\begin{figure}
\begin{center}
\includegraphics[width=7.5cm]{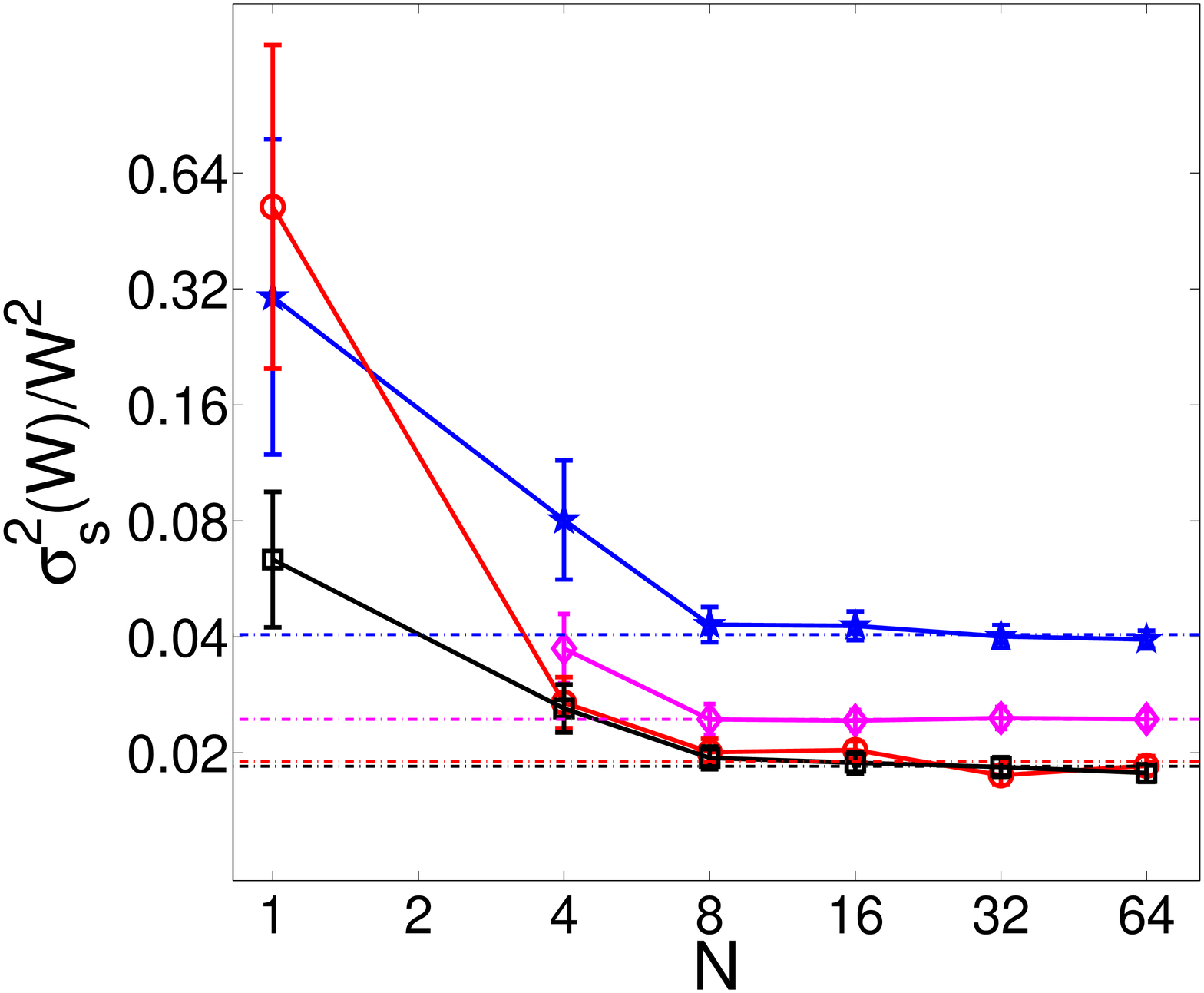}
\includegraphics[width=7.5cm]{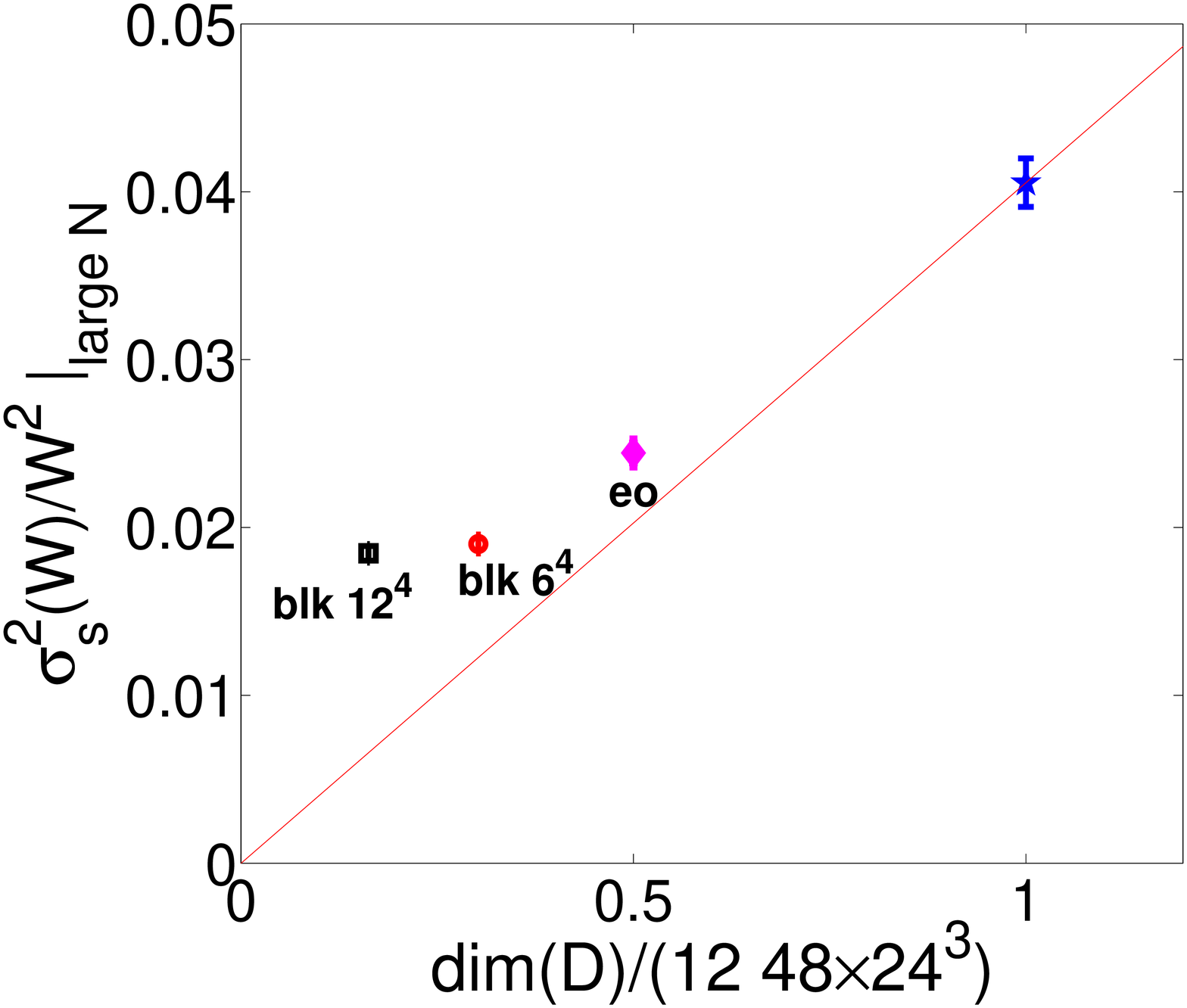}
\end{center}
 \vspace{-0.5cm}
 \caption{\small The figures show the stochastic fluctuations in the case of two flavor mass reweighting
	  by using mass interpolation and domain decomposition for one configuration of the $48\times24^3$ ensemble.
	  In the left figure we analyze the scaling in the number of mass interpolation steps $N$ while
	  the total number of global inversions $N\cdot N_{hit}$ is fixed to $640$. We plot the estimated relative
	  stochastic variance $\sigma_s^2(W)/W^2$ against the number of interpolation steps $N$. The analysis shows
	  that for $N\geq8$ it makes no difference if one increases $N_{hit}$ or $N$. If the eigenvalues of $M_i^\dagger M_i$
	  are close enough to one $\sigma_s^2(W)/W^2$ scales with $1/N N_{hit}$. We fit
	  the asymtotic plateau for the total operator (star,blue), the even-odd preconditioned operator
	  (equivalent to a Schur complement with blocks of length l=1) (diamonds,magenta),
	  the Schur complement with $6^4$-blocks (circle, red) and with $12^4$ blocks (square,black).
	  The right figure shows the results of the plateau fit against the total dimension of the operators divided
	  by the dimension of the global Wilson-Dirac operator. For the Schur complement we only take
	  the dimension of the projector $P$ into account. 
}
 \label{fig:twofl}
\end{figure}

In addition Eq. (\ref{eq:evM}) implies that a smaller 
mass-shift would shift the eigenvalues closer to one. 
This is easily achieved if one introduces an interpolation in the mass and 
by splitting up the ratio matrix in several ratio matrices 
$D(m_1)/D(m_2) = \left\{D(m_1)/ D(m_i)\right\} \left\{ D(m_i)/ D(m_2) \right\}$.
This technique works only if no real eigenvalue of the Wilson-Dirac operator becomes negative.
In this case the ratio matrix gets eigenvalues which are smaller than 1/2
and the stochastic estimation fails. If this happens one has to use additional methods,
like exact eigenvalue calculation, to calculate the reweighting factor in an appropriate
way.
The estimation of the reweighting factor with the domain decomposition is now given by
\begin{equation}
  W =W_{gl} \cdot \prod_{k=1}^{N_{blk}} \frac{\det D^2_k(m_2)}{\det D^2_k(m_1)}
\end{equation}
where $k$ labels the white and black blocks. For moderate block sizes $l^4\leq6^4$ the exact calculation of the block determinants is feasible
while the global factor $W_{gl}$ is estimated by
using $N$ mass interpolation steps and $N_{hit}$ estimations of each ratio
\begin{equation}
  W_{gl} = \prod_{i=1}^{N} \left\{ \frac{1}{N_{hit}} \sum_{j=1}^{N_{hit}} e^{-\eta_{i,j}^\dagger (\hat{M}_{i}^\dagger \hat{M}_{i} - 1 )\eta_{i,j}} \right\}
  \label{eq:Wgl}
\end{equation}
where the $i$th ratio matrix is given by $M_i=\hat{D}^{-1}(m_i) \hat{D}(m_{i-1})$ with
the Schur complement $\hat{D}(m_i)$ depending on
the $i$th mass \mbox{$m_i = i/N \cdot m_2 + (N-i)/N \cdot m_1$}.
Inverting the Schur complement costs one inversion of the Dirac operator.

In practice it is now easy to control the stochastic fluctuations
by changing the number of inversions $N\cdot N_{hit}$, this is possible
as long as there is no zero-crossing of the eigenvalues of $D(m)$.
To avoid a wrong estimation (zero-crossings for $m_1>m_2$) it is necessary to control the variance of
each factor in (\ref{eq:Wgl}), which can be estimated by setting $N_{hit}\geq 6$. 
Increasing $N$ or $N_{hit}$ is comparable, if the eigenvalue distribution of the ratio matrix $M^\dagger M$ is close to
one, which can be achieved by increasing $N$ to sufficient value (see Fig. (\ref{fig:twofl})).

Fig. (\ref{fig:twofl}) also shows the effect of using the Schur complement instead of the total Dirac operator.
For block sizes $l\geq1$ it is two times more efficient to use the Schur complement.
Another conclusion is that stochastic fluctuations do not scale with the dimension of the
operator. It is obvious that the remaining IR-modes dominate the fluctuations.
In general we find that the stochastic fluctuations scale with $\Delta m^2 V /(N\cdot N_{hit})$.

\subsection{Ensemble Fluctuations}

\begin{figure}
\begin{center}
\includegraphics[width=7.5cm]{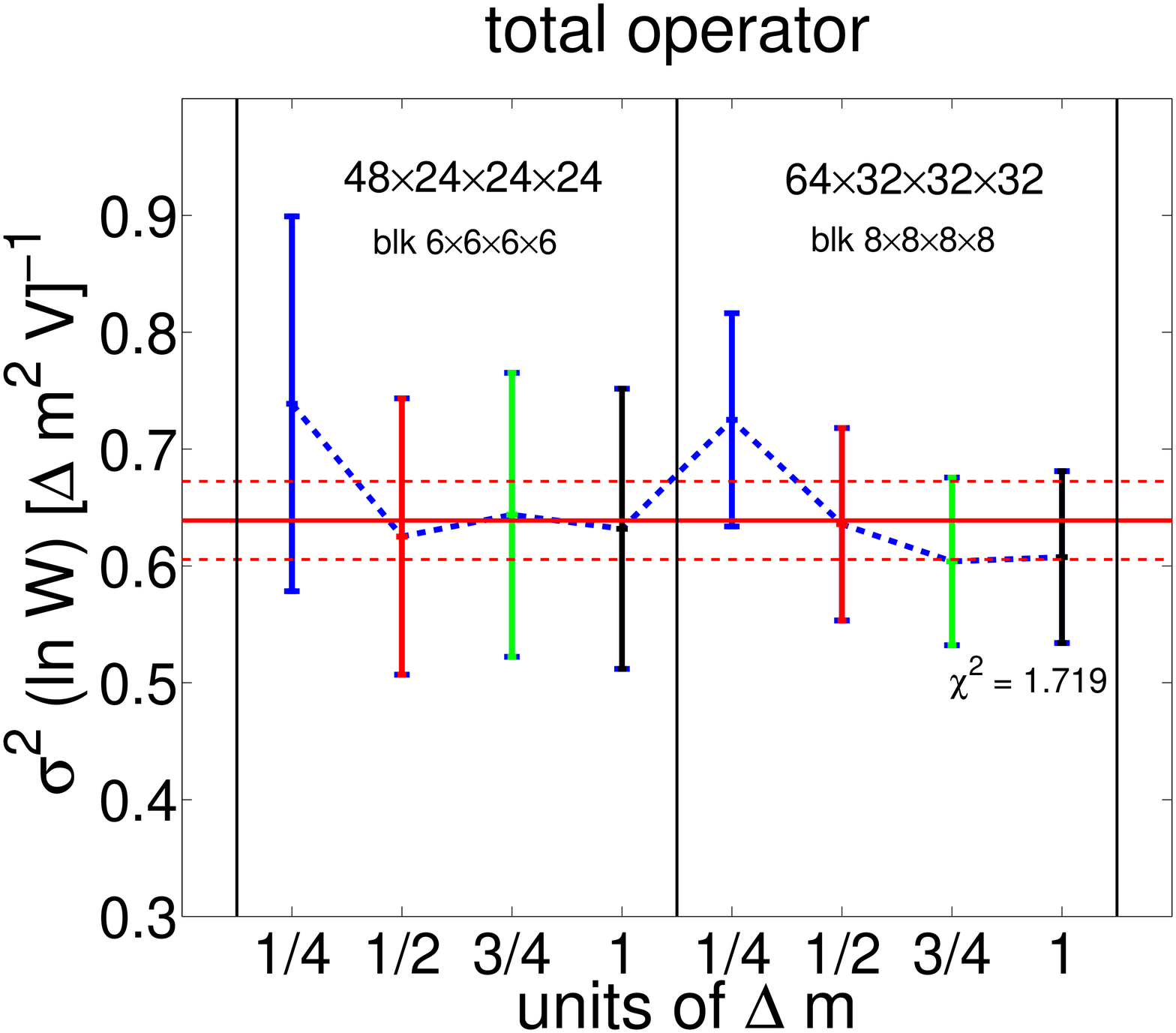}
\includegraphics[width=7.5cm]{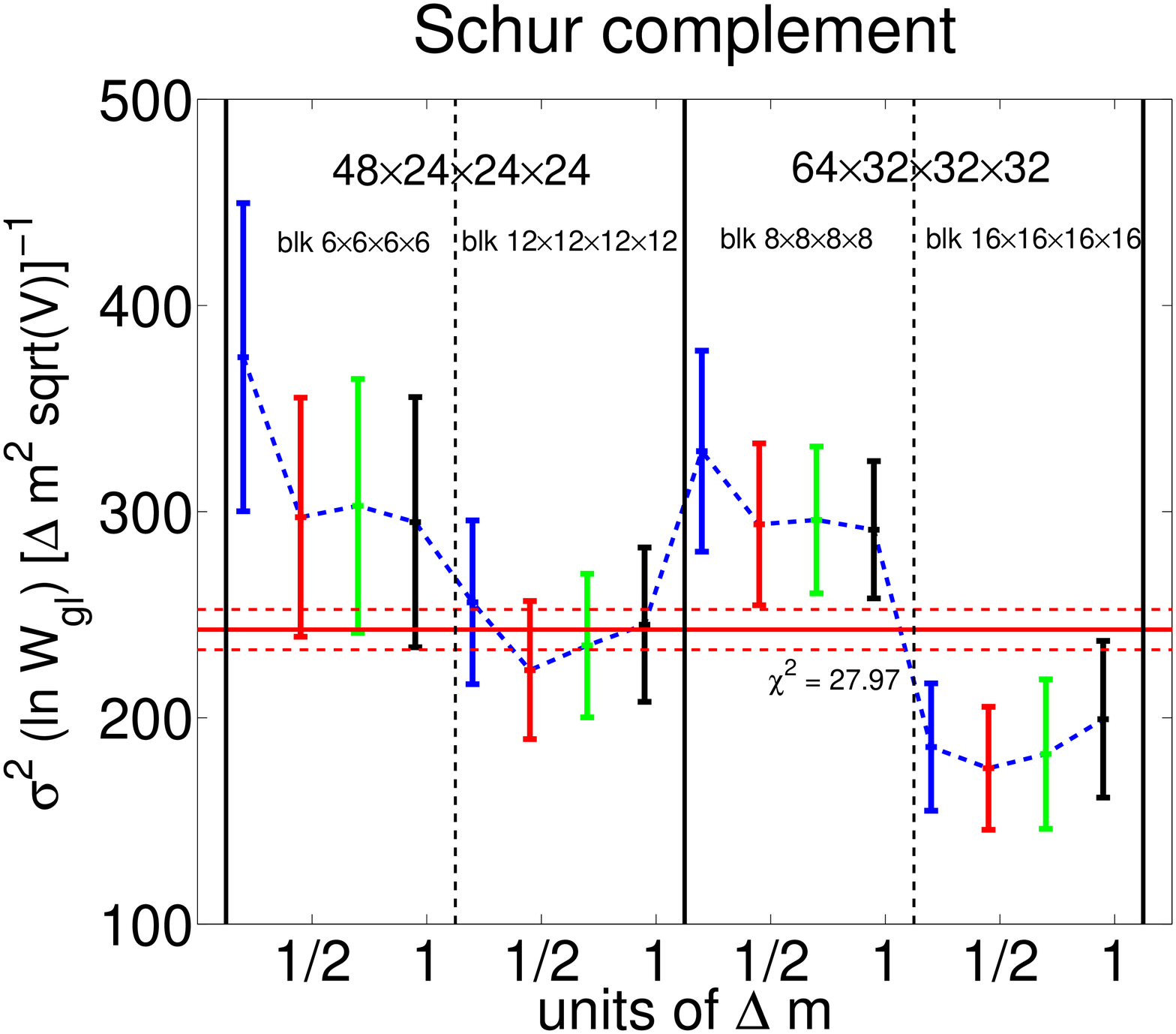}
\end{center}
 \vspace{-0.5cm}
 \caption{\small The figures show the scaling of $\sigma^2(\ln W)$ with the volume $V$ and the mass shift $\Delta m$ by using 60-100 configurations. 
The left figure shows the fluctuations of the global factor $\sigma^2(\ln W)$ multiplied by $1/V \Delta m^2$ against several
mass shifts $\Delta m$ while on the left side we plot the small volume and on the right side the bigger one (we write the parameters
for the bigger in brackets). The ensemble noise is estimated with $N_{hit} = 6$, $N=16,(48)$ and domain decomposition with $6^4(8^4)$ blocks.
The $8^4$ blocks are decomposed further in a $8^4$ Schur complement with Dirichlet boundaries and $4^4$ blocks.
The right figure show the fluctuations for the global Schur complement multiplied by $1/\sqrt{V} \Delta m^2$ against several mass shits and 
different Schur complements. The constant fit illustrate the weak volume dependence while $\sqrt{V}$ is an upper estimate for this dependence.}
 \label{fig:enfl}
\end{figure}

The ensemble fluctuations enter the game if one wants to calculate an observable,
which is given for the target mass by  $\langle\mathrm{O}\rangle_{m_2} =  \frac{\langle\mathrm{O} W\rangle_{m_1}}{\langle W\rangle_{m_1}}$.
The total variance of such an observable gets the form \cite{Liu:2012gm},\cite{Aoki:2010dy} 
  
\begin{equation}
 {{\rm{var}}( \mathrm{O} )}/{N_{cnfg}}\sim \frac{\delta \mathrm{O}^2}{N_{cnfg}} \tau_{corr} \left(\frac{{\rm{var}}(W)}{\langle W \rangle^2} + 1\right)
\end{equation}
where $\delta \mathrm{O}^2$ is the variance of the observable without the reweighting factor, $N_{cnfg}$ the total number of configurations and
$\tau_{corr}$ the autocorrelation time.
We observe that the mass reweighting factor is
distributed like a log-normal distribution 
$\rho( W ) \sim \frac{1}{W} \exp\left\{- \frac{(\ln W - \mu)^2}{2\sigma^2}\right\}$.
Then it is straightforward to show that the ensemble noise is given by
\begin{equation}
 \left(\frac{{\rm{var}}(W)}{\langle W\rangle^2} + 1\right) =  e^{\sigma^2}
\end{equation}
with $\sigma^2 = {\rm{var}}(\ln W)$.

In order to study the scaling of $\sigma^2$, we fix the stochastic noise to a small and volume independent value.
From Fig. (\ref{fig:enfl}) it follows directly that $\sigma^2 = k_1 \cdot \Delta m^2 \cdot V$ for some constant $k_1$, while we observe that the Schur complement
has only a weak $V$ dependence. We appraise it with $\sqrt{V}$. 
 The $V$ dependence of $W$ emerges through a
large correlation between the factors of the block operators and the Schur complement.
In general mass reweighting in large volume is limited to small values of $\Delta m$.

\section{One Flavor Mass Reweighting}

\begin{figure}
\vspace{-0.5cm}
\begin{center}	
\includegraphics[width=7.5cm]{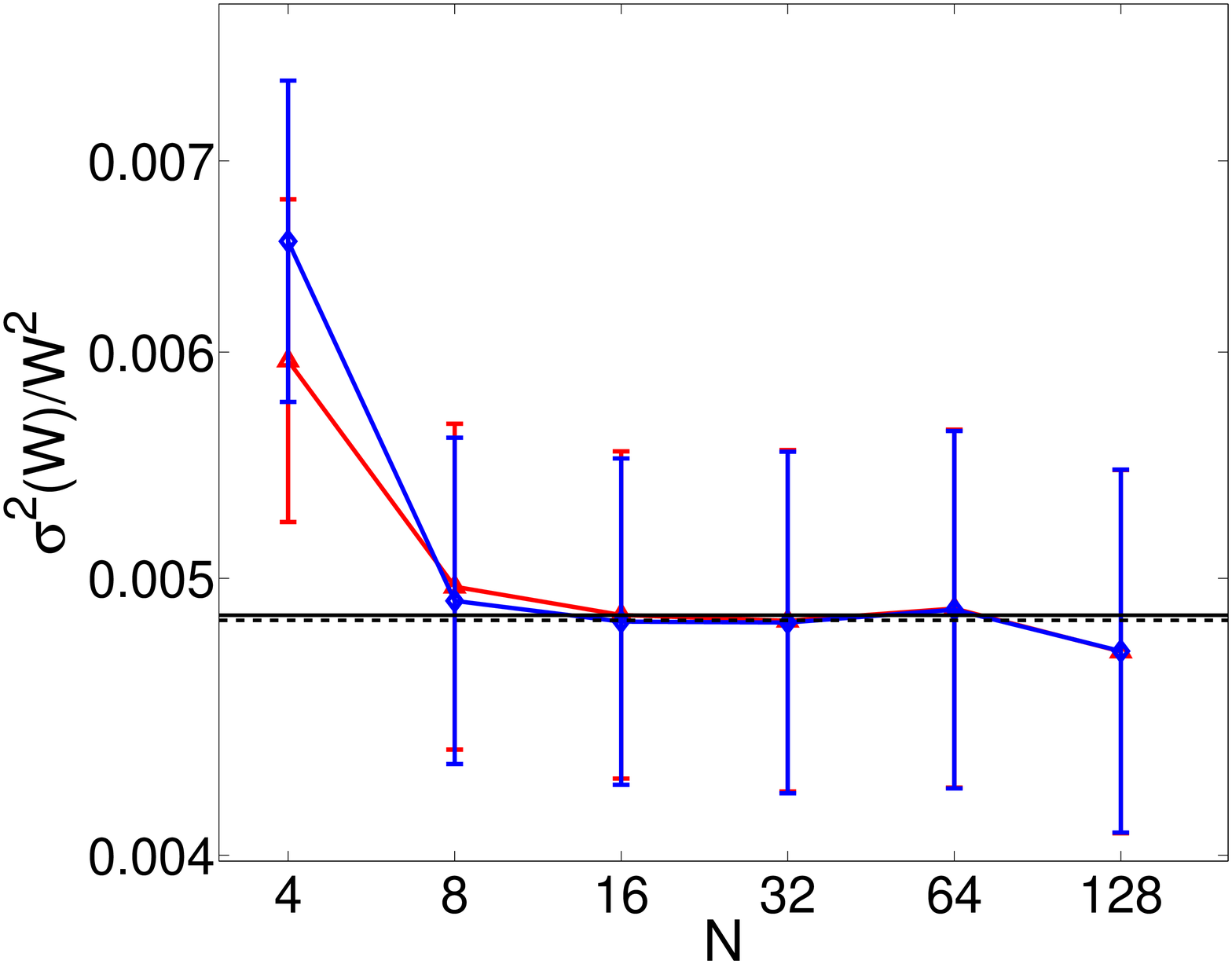}
\includegraphics[width=7.5cm]{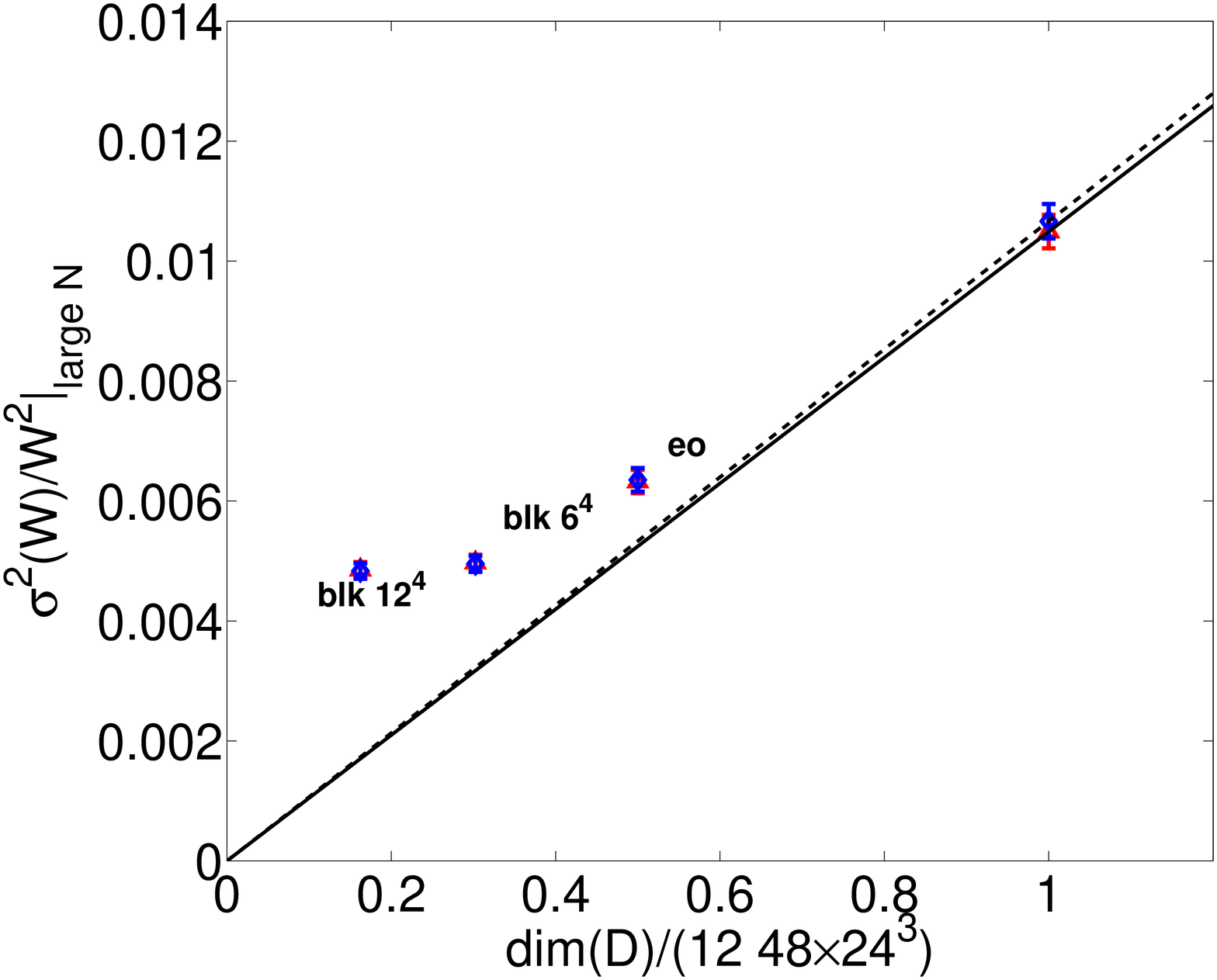}
\end{center}
 \vspace{-0.5cm}
  \caption{\small The figures show the scaling of the stochastic fluctuations in the case of one flavor mass reweighting 
  for one configuration of the smaller volume like in Fig. (\protect\ref{fig:twofl}). We compare the one flavor reweighting with
  the root-trick \protect\cite{Aoki:2012st}. The root-trick is a biased estimator by using the square root of the two-flavor estimation 
  $1/\det M = \sqrt{\int \rm{D}[\eta] \rm{D}[\eta^\dagger] \exp\left\{-\eta^\dagger M^\dagger M \eta\right\} }
  \rightarrow \sqrt{1/N_{hit} \sum\nolimits_{i=1}^{N_{hit}} \exp\{-\eta_i^\dagger(M^\dagger M - 1)\eta_i\}}$.
  The right figure shows the estimated relative variance $\sigma^2(W)/W^2$ of the global Schur complement
  with $12^4$ blocks against the number of mass interpolation steps $N$. The total number of inversions
  for each point is constant with $N\cdot N_{hit} = 640$. The plot shows the difference of the stochastic 
  fluctuations for one flavor case Eq. (\protect\ref{eq:ofm}) (red,triangle) and the root-trick (blue,diamonds). One can see that
  for all points the variance is finite, the one flavor integral exists. The right figure shows the constant fit
  to the $1/N N_{hit}$-plateau for several operators (compare Fig. (\protect\ref{fig:twofl})).
}
 \label{fig:1fl}
\end{figure}

Nature motivates one flavor reweighting. There are many effects which depend only on the specific
quark, like isospin splitting of the up- and down-quark. Also for corrections of a not exactly tuned strange quark mass it is
necessary to calculate the mass reweighting factor for one flavor. For that we introduce the integral
\begin{equation}
\frac{1}{\det M} = \int {\rm D}[\eta] {\rm D}[\eta^\dagger] \exp\left\{-\eta^\dagger M \eta\right\} \rightarrow \frac{1}{N_{hit}} \sum_{i=1}^{N_{hit}} e^{-\eta_i^\dagger(M - 1)\eta_i}
\label{eq:ofm}
\end{equation}
which is well defined only if ${\rm Re} (\lambda(M)) > 0$.
The variance of the stochastic estimation is given by
\begin{equation}
 \sigma^2_s = \frac{1}{\det (M^\dagger + M -1)} - \frac{1}{\det M^\dagger M}
\end{equation}
which is defined if $\lambda (M^\dagger + M) > 1 $. So it is possible to estimate the reweighting factor
$1/\det M$ as long as the variance is defined. 
In general the scaling of one flavor mass reweighting is comparable with the two flavor case (see Fig. (\ref{fig:1fl})),
but obviously there are some differences. The estimate is complex. In practice one can use this
to improve the estimator: because of the $\gamma_5$ Hermiticity the expectation value is real and 
one can neglect the imaginary part.
We found that the errors are comparable to the square root trick \cite{Aoki:2012st}. 
In the case that a real eigenvalue
becomes negative it is not possible by using mass interpolation to ensure that the integral is defined.
Another issue is that the estimate being complex could have a negative sign. In this case the one flavor estimate
in eq. (\ref{eq:ofm}) works (provided the integral is defined) but not the square root trick. We do not detect such problem
if we suppress the stochastic noise to a proper level which is easily achieved by increasing
the number of mass interpolation steps.
The proposed one flavor estimation has many advantages and
should be used in future applications.

\section{Conclusion}

In these proceedings we analyze the scaling behavior of mass reweighting by studying the stochastic and ensemble fluctuations
with the methods mass interpolation and domain decomposition. We find that the stochastic fluctuations
scale like $\Delta m^2 V /(N\cdot N_{hit})$. By using domain decomposition with block sizes with $l\geq 1$
the fluctuations are reduced at least by a factor two. The ensemble fluctuations of the full operator scales like $\Delta m^2 V$
while for the Schur complement the volume dependence is weaker and compatible with $\Delta m^2 \sqrt{V}$. 

Assuming that $ \sigma_{tot}^2 (\ln W) = \sigma^2(\ln W) + \frac{\sigma_s^2}{W^2}$
the cost for the mass reweighting of the total operator can be deduced from the number of the original
configurations needed, given by
\begin{equation}
\frac{N_{cnfg}}{\tau_{corr}}  = N_{eff} \cdot \exp\left\{\Delta m^2 \cdot V \left(k_1 + \frac{k_2}{N N_{hit}}\right)\right\} + \mathcal{O}(\Delta m^3) 
\label{eq:cost}
\end{equation}
for constants $k_1$ and $k_2$ which depends on the ensemble parameters.
Here we use the definition of the number of effective configuration $N_{eff} = \frac{N_{cnfg}}{\tau_{corr}}\left({\rm{var}}(W)/\langle W\rangle^2 + 1\right)$ 
of \cite{Liu:2012gm} and an analytic expansion of Eq. (\ref{eq:vst}).
For $N_{eff} = 50$ we get \mbox{$N_{cnfg}/\tau_{corr} = 1517 ( 241509$ (for the bigger volume) )} for $N\cdot N_{hit} = 32 $ by fixing $k_2 /( k_1 N_{hit} N ) = 0.11$.
If one consider a reweighting range of $\Delta m/2$ the numbers change to $N_{cnfg}/\tau_{corr}= 117 (741)$ using the same numbers of inversions $N\cdot N_{hit} = 32 $.
The total cost of evaluating the reweighting factor only scales with the volume $V$ but through the $V$ dependence of the
ensemble fluctuations mass reweighting becomes rapidly inefficient for larger volumes.
This limits the reweighting range in $m$. 

To conclude there are many more details to discuss and to describe in a more general style in the framework 
of mass reweighting, like an analytic formula to characterize the stochastic estimation, estimation
with zero crossings or a proof for the one flavor integral Eq. (\ref{eq:ofm}). We want to address this soon in an adequate way.


\begin{thebibliography}{99}
{
\bibitem{Finkenrath:2012az} 
  J.~Finkenrath, F.~Knechtli and B.~Leder,
  arXiv:1204.1306 [hep-lat].
\bibitem{Hasenfratz:2008fg}
  A.~Hasenfratz, R.~Hoffmann and S.~Schaefer,
  Phys.\ Rev.\ D {\bf 78}, 014515 (2008) \newline
  [arXiv:0805.2369 [hep-lat]].
\bibitem{Luscher:2005rx} 
  M.~L\"uscher,
  Comput.\ Phys.\ Commun.\  {\bf 165}, 199 (2005)
  [arXiv:0409106 [hep-lat]].
\bibitem{Ferrenberg:1988yz} 
  A.~M.~Ferrenberg and R.~H.~Swendsen,
  Phys.\ Rev.\ Lett.\  {\bf 61}, 2635 (1988).
\bibitem{Fritzsch:2012wq}
  P.~Fritzsch, F.~Knechtli, B.~Leder, M.~Marinkovic, S.~Schaefer, R.~Sommer and F.~Virotta, \newline
  Nucl.\ Phys.\ B {\bf 865}, 397 (2012)
  [arXiv:1205.5380 [hep-lat]].
\bibitem{Hasenfratz:2002ym}
  A.~Hasenfratz and A.~Alexandru,
  Phys.\ Rev.\ D {\bf 65}, 114506 (2002)
  [hep-lat/0203026].
\bibitem{Liu:2012gm}
  Q.~Liu, N.~H.~Christ and C.~Jung,
  arXiv:1206.0080 [hep-lat].
\bibitem{Aoki:2010dy} 
  Y.~Aoki {\it et al.}  [RBC and UKQCD Collaborations],
  Phys.\ Rev.\ D {\bf 83}, 074508 (2011)\newline
  [arXiv:1011.0892 [hep-lat]].
\bibitem{Aoki:2012st} 
  S.~Aoki, K.~I.~Ishikawa, N.~Ishizuka, K.~Kanaya, Y.~Kuramashi, Y.~Nakamura, Y.~Namekawa and M.~Okawa {\it et al.},
  Phys.\ Rev.\ D {\bf 86}, 034507 (2012)
  [arXiv:1205.2961 [hep-lat]].
}
\end{thebibliography}
\end{document}